\documentclass[final,3p,times,twocolumn]{elsarticle}
 \biboptions{comma,sort&compress}
\usepackage{here}
 \usepackage{graphicx}
  \usepackage{epsfig}

\def\beq{\begin{equation}}
\def\eeq{\end{equation}}
\def\bea{\begin{eqnarray}}
\def\eea{\end{eqnarray}}

\newcommand{\bbbar}     {\ensuremath{\mathrm{b\bar{b}}}}
\newcommand{\epem}              {\ensuremath{\mathrm{e^+e^-}}}

\newcommand{\as}                {\ensuremath{\alpha_\mathrm{S}}}

\newcommand{\asmz}              {\ensuremath{\alpha_\mathrm{S}(\mathrm{M}_{\mathrm{Z^0}})}}

\newcommand{\mz}                {\ensuremath{\mathrm{M}_{\mathrm{Z^0}}}}

\newcommand{\chisqd}    {\ensuremath{\chi^2/\mathrm{d.o.f.}}}
\newcommand{\xmu}               {\ensuremath{x_{\mu}}}

\newcommand{\ycut}              {\ensuremath{y_{\mathrm{cut}}}}

\newcommand{\stat}              {\ensuremath{\mathrm{(stat.)}}}

\newcommand{\expt}               {\ensuremath{\mathrm{(exp.)}}}
\newcommand{\had}               {\ensuremath{\mathrm{(had.)}}}
\newcommand{\theo}              {\ensuremath{\mathrm{(theo.)}}}

\newcommand{\yy}                {\ensuremath{y_{23}}}

\newcommand{\rs}                {\ensuremath{\sqrt{s}}}

\newcommand{\invpb}     {\ensuremath{\mathrm{pb}^{-1}}}

\newcommand{\py}                {PYTHIA}
\newcommand{\hw}                {HERWIG}
\newcommand{\ar}                {ARIADNE}

\newcommand{\result} {\ensuremath{\asmz=0.1199\pm0.0010\stat\pm0.0021\expt\pm0.0054\had\pm0.0007\theo}}

\newcommand{\resultxmuopt} {\ensuremath{\asmz=0.1204\pm0.0009\stat\pm0.0021\expt\pm0.0059\had\pm0.0008\theo}}

\journal{Nuc. Phys. (Proc. Suppl.)}

\begin{document}

\begin{frontmatter}



\title{Measurement of the strong coupling \as\  in \epem-annihilation \\ using the three-jet rate}

 \author[label1]{Jochen Schieck\corref{cor1}}
  \address[label1]{Ludwig-Maximilians-Universit\"at M\"unchen, Am Coulombwall 1, D-85748 Garching, Germany \\ and
  Excellence Cluster Universe, Boltzmannstr. 2, D-85748 Garching, Germany.}
\cortext[cor1]{Speaker}
\ead{Jochen.Schieck@lmu.de}



\begin{abstract}
\noindent
We present a measurement of the strong coupling \as\ using data collected with the JADE detector
at centre-of-mass energies between 14 and 44 GeV. The three-jet rate as a function
of the transition parameter \ycut\ is determined using the Durham jet algorithm and the 
distribution is compared to QCD predictions. Recent theoretical calculations predict the 
three-jet rate at next-to-next-to-leading order. For the first time a measurement of \as\ is
presented using QCD predictions at next-to-next-to-leading order matched to predictions at next-to-leading 
logarithmic approximation, with subleading terms being included as well. We obtain  \\
\result, \\
being consistent with the world average value of \as.
\end{abstract}
\end{frontmatter}
\section{Introduction}
\label{Intro}
The strong coupling \as\ is one of the fundamental parameters of the Standard Model of particle physics. Several possibilities
exist to determine this parameter, one method being presented in this paper. In \epem-annihilation with quarks in the final
state hard gluons can be emitted from the quark with the emission probability being proportional to the strong coupling \as. Events
with a hard gluon are identified by observing at least three particle jets in the detector.
The theory of strong interaction, Quantum Chromo Dynamics (QCD), allows to predict the rate of three-jet events as a function
of a single parameter, the strong coupling \as. Recent theoretical 
improvements~\cite{GehrmannDeRidder:2007jk,GehrmannDeRidder:2007hr,GehrmannDeRidder:2008ug,Weinzierl:2008iv}
allow to calculate the three-jet rate at next-to-next-to-leading-oder (NNLO)
and a first measurement with data collected by the ALEPH detector at 91~GeV using the three-jet rate at a single 
resolution parameter was published~\cite{Dissertori:2009qa}. \par
In this paper we present the measurement of \as\ using data collected with the JADE detector between the years 1979 and 1986.
For the first time NNLO predictions are matched with next-to-leading logarithmic approximations (NLLA), 
including subleading soft logarithms, to measure \as\ with the three-jet rate. 
The complete description of the analysis can be found in~\cite{Schieck:2012mp}.
\section{The three-jet rate determined with the Durham clustering scheme}
\begin{figure}[hbt] 
{\epsfig{figure=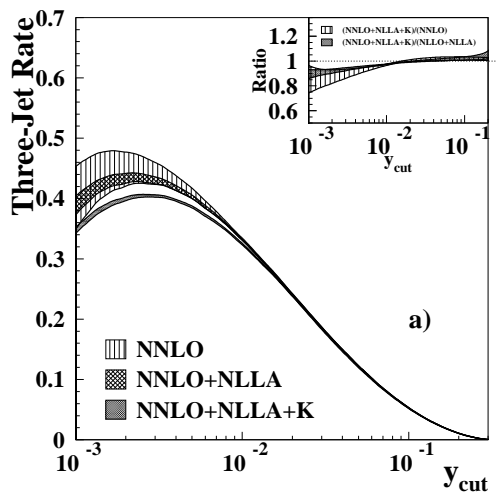,height=70mm}}
\caption{\scriptsize QCD predictions for partons clustered to three jets using the Durham algorithm evaluated at 
a centre-of-mass energy of 35~GeV and 
the strong coupling being set to \asmz=0.1180. The predictions do not contain corrections expected from hadronisation effects. The figure
shows the fixed order prediction (NNLO) together with the fixed order prediction matched to NLLA (NNLO+NLLA) and the
matched predictions taking subleading soft logarithms into account (NNLO+NLLA+K). The band reflects the theoretical
uncertainty estimated by setting the renormalisation scale factor \xmu\ to 0.5 and 2.}
\label{FigNNLO} 
\end{figure}

In order to count the number of jets in the final state the reconstructed particle tracks and neutral clusters have to be 
clustered to jets. Several clustering schemes are available and we use the Durham clustering algorithm~\cite{Catani:1991hj}.
The number of jets depends on the resolution parameter \ycut, a measure indicating the separation between the second and
the third reconstructed jet. The fixed order prediction of the three-jet rate determined with the Durham clustering scheme is matched to 
NLLA predictions using the $R$-matching scheme. Parts of the subleading soft logarithms, which are not included
in the NLLA calculations, can be systematically controlled and included by adding a renormalisation scheme 
dependent $K$-term~\cite{Nagy:1998kw}. Including the $K$-term significantly improves the description of the data.
QCD predictions of the three-jet rate, without the expected contribution from hadronisation effects, are shown 
in Fig.~\ref{FigNNLO}.
\section{The JADE Detector, Data and Monte Carlo sample}
\begin{figure}[hbt] 
{\epsfig{figure=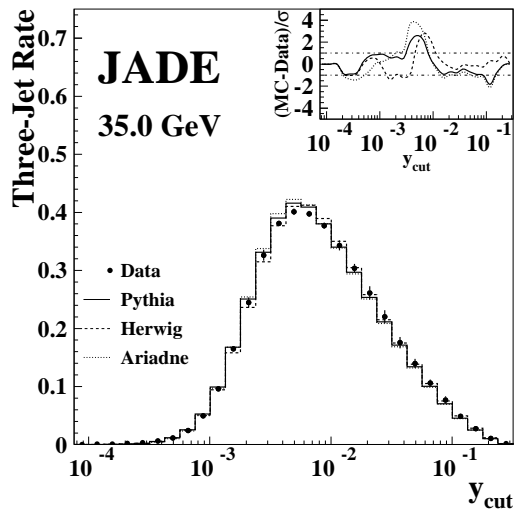,height=70mm}}
\caption{\scriptsize The three-jet rate as a function of the resolution parameter \ycut\ for data taken at a centre-of-mass energy of  35~GeV. 
The histogram shows the three-jet rate in comparison to simulation using \py,\hw\ and \ar\ Monte Carlos models. The 
insert shows the difference to the model normalized to the combined statistical and systematic uncertainty.}
\label{figMCData} 
\end{figure}

The JADE detector collected data in \epem-annihilation at the PETRA accelerator at DESY between 1979 and 1986. 
We use data taken at 14~GeV, 22~GeV, 34.6~GeV, 35~GeV, 
38.3~GeV and  43.8~GeV, adding up to a total integrated luminosity of  about 195 \invpb. The number of selected hadronic
events ranges between about 1000 events at 14~GeV and 20000 events at 35~GeV. The selection of hadronic events 
is based on the event multiplicity, momentum imbalance and on the total visible energy.
The events are corrected for acceptance effects and the QCD predictions for the hadronisation 
of partons to hadrons. For these correction procedures Monte Carlo (MC) 
samples generated with \py, \hw\ or \ar\ are used, with the generators being  
tuned to events taken with the OPAL detector at LEP at a centre-of-mass energy of 91~GeV. 
The  expected contribution from $\epem\to\bbbar$-events, as predicted by simulation,
is subtracted from the observed three-jet rate. The excellent description
of the three-jet event rate by  simulation is visualised in Fig.~\ref{figMCData}. 
\section{Measurement of the strong coupling \boldmath{\as}}
\begin{figure}[hbt] 
{\epsfig{figure=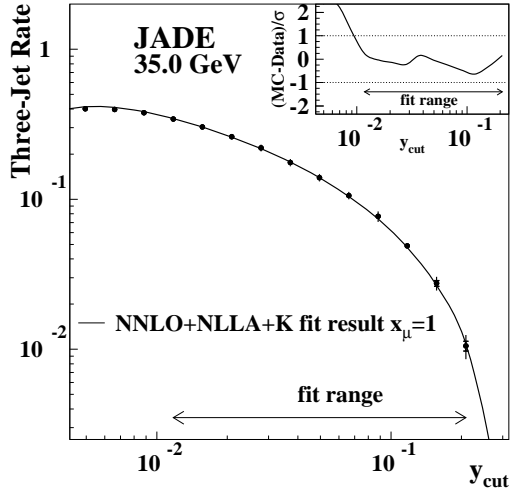 ,height=70mm}}
\caption{\scriptsize Result of the fit from data taken at a centre-of-mass energy of 35~GeV. The insert reflects the difference
between the QCD prediction calculated with the strong coupling returned from the fit and the measured distribution, 
normalized to the combined statistical and experimental uncertainty.}
\label{FigFit} 
\end{figure}

The measured and corrected three-jet rate together with the matched NNLO+NLLA+K QCD predictions is
used to determine the strong coupling \as. A $\chi^{2}$-value, obtained from the difference between
the measured three-jet rate and the QCD predictions,  is calculated and minimised. The QCD predictions are applied with the
renormalisation scale factor set to the natural choice \xmu=1. Events simulated with the \py\ event generator are used
to correct the predicted three-jet rate for hadronisation effects. The fit range is determined by requiring the correction from hadronisation
and detector effects to be small. In addition the leading log contribution in the QCD prediction is required to be well 
below unity. The correlations between the different \ycut\ bins are determined using MC events.
The result obtained at a centre-of-mass energy of 35~GeV from the fit to the three-jet rate is shown together with 
the three-jet rate distribution in Fig.~\ref{FigFit}. 
\subsection{Systematic Uncertainties}
Several sources of systematic uncertainties are considered for the measurement of the strong coupling \as. 
The fit is repeated and the difference between the variation and the
default analysis is taken as the symmetric systematic uncertainty. The different sources of systematic uncertainties 
are arranged in three categories: 
\begin{itemize}
\item {\bf Experimental uncertainties:}
The fit is repeated with a somewhat modified event selection, different reconstruction software, different correction for 
detector effects or a slight variation of the fit range. The systematic uncertainties are evaluated applying these 
variations and added in quadrature. The main contribution comes from using different detector correction procedures and 
using different reconstruction software.
\item {\bf Hadronisation uncertainties:}
Different MC models are applied to adjust the QCD predictions for hadronisation effects. The largest systematic
uncertainty comes from evaluating the hadronisation correction with the \hw\ event generator instead of the 
\py\ event generator, as used in the default method. 
\item {\bf Theoretical uncertainties:}
In the default fit the renormalisation scale parameter \xmu = $\mu/\rs$ is set  to the natural choise \xmu=1. As a systematic
variation \xmu\ is set to 0.5 and 2 and the larger difference to the default fit is taken as the theoretical systematic uncertainty.
\end{itemize}
\subsection{Results}
\begin{figure}[hbt] 
{\epsfig{figure=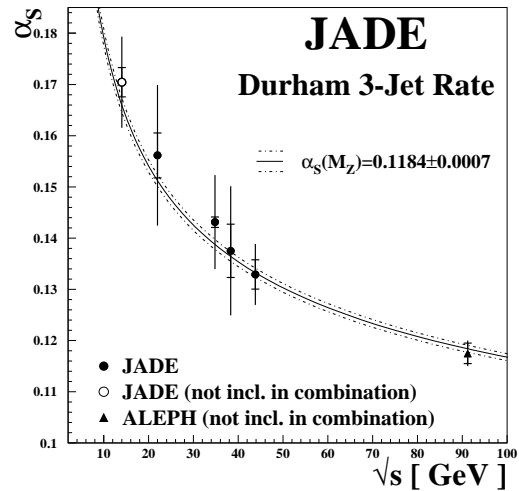,height=70mm}}
\caption{\scriptsize Measurements of the strong coupling \as\ determined with 
the three-jet rate using the Durham jet-finder algorithm and NNLO order predictions.
For the measurements between 14 and 44~GeV the QCD predictions are matched
with NLLA including soft logarithms. The energy points at 34.6 and 35~GeV are combined to a single entry.
The line corresponds to the world average value with the corresponding uncertainty~\cite{Bethke:2009jm}.  }
\label{FigRunAlphaS} 
\end{figure}

The measurement of \as\ is performed at each energy point separately and the final result is obtained by combining the
results of \as\ at the various centre-of-mass energies. Due to large hadronisation corrections the result from the 
fit to the data taken at 14~GeV is not included in the combination. The combined value for \as\ evolved to the energy \mz\ is 
\result, consistent with the world average value of $0.1184\pm0.0007$~\cite{Bethke:2009jm}. 
The measurements of \as\ at the various energy points is shown in Fig.~\ref{FigRunAlphaS}. The evolution of \as\ measured
with the three-jet rate as a function of the centre-of-mass energy is consistent with the expectation from QCD.
\subsection{Study of the renormalisation scale dependence}
Besides fixing the renormalisation scale \xmu\ to the natural choice the
fit is repeated with \xmu\ and \as\ being free in the fit. The renormalisation
scale obtained by the fit is consistently below unity, but with large statistical
uncertainties and being consistent with one. The large uncertainty 
can be understood by looking in more detail to the variation of \as\ with respect to
\xmu. The change in \as\ is rather flat, leading to small theoretical uncertainties
and to large statistical uncertainties for the renormalisation scale \xmu. 
The results obtained from the various energy points are combined to a single 
value, leading to \resultxmuopt, again being consistent with the world average value.
\subsection{Measurement of \boldmath{\as} using fixed order predictions}
The fit is repeated with fixed order predictions only and \xmu\ being set to one 
or leaving \xmu\ as free parameter in the fit. For the fit with 
\xmu\ being set to the natural choice the \chisqd\ is considerably worse than the 
one obtained with matched QCD predictions (almost a factor four in the \chisqd\ for the
fit to the data collected at 35~GeV). This points to missing higher order
terms in the QCD prediction.
In addition the fit result also shows 
an increased sensitivity to the fit range. Using matched
NNLO+NLLA+K QCD predictions the dependency on the fit range is almost 
negligible, while for NNLO predictions the choice of the lower \ycut\ value 
in the fit range has a significant impact on the result. \par
Leaving \as\ as well as \xmu\ free in the fit leads to a significantly improved 
description of the fit with \chisqd\ values similar to the one obtained from the fit with 
matched QCD predictions. However, the value obtained
for the scale \xmu\ is rather low (around 0.2) and within the statistical uncertainty
inconsistent with being one. The uncertainty on the fitted scale parameter \xmu\ 
is about one
order of magnitude smaller than the uncertainty on the scale parameter obtained
with matched predictions. This indicates that the sensitivity of \as\ with respect
to the renormalisation scale \xmu\ is increased compared to the fit with matched 
QCD predictions. 
\section{Conclusions}
We present the first measurement of \as\ determined with the Durham three-jet rate 
using matched NNLO+NLLA+K QCD predictions.  For this measurement hadronic 
events from \epem-annihilation taken at centre-of-mass energies between 
14 and 44~GeV are used. The result 
\result\
is consistent with the world average value. The largest uncertainty
originates from modeling the transition from parton level, as predicted from the 
perturbative QCD calculations, to the hadron level, as observed by the experiment. 
The increased precision of perturbative
QCD predictions, together with large hadronic \epem\ event samples, will allow to scrutinize  
MC generators modeling the transition from partons to hadrons. A measurements of \as\ 
using fixed order predictions only cannot describe the data satisfactorily. A fit
leaving \as\ as well as \xmu\ free returns a good description of the data, but the
fitted value of \xmu\ is inconsistent with one. \par
\begin{figure}[hbt] 
{\epsfig{figure=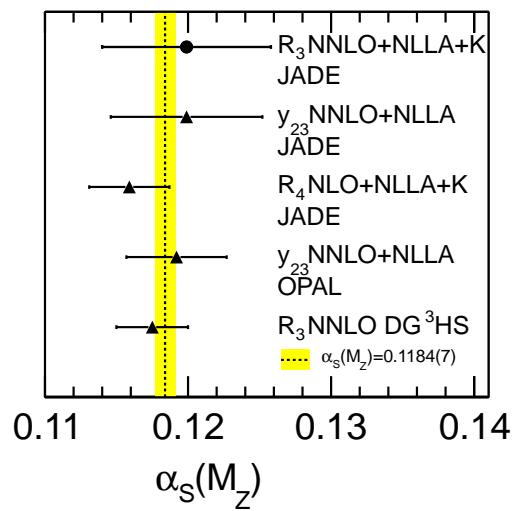,height=70mm}}
\caption{\scriptsize Summary of \as\ measurements using the three-jet rate, the four-jet rate and the 
differential \yy\ distribution applied to \epem-data taken between 14 and 91~GeV. The first line
corresponds to the measurement summarized in this paper, the second line to~\cite{Bethke:2008hf}, the 
third line to~\cite{Schieck:2006tc}, the fourth line to~\cite{Abbiendi:2011nnlo} and the last line to~\cite{Dissertori:2009qa}. 
The vertical band indicates the world average value of \as~\cite{Bethke:2009jm}.}
\label{FigSummary} 
\end{figure}

Fig.~\ref{FigSummary} summarizes the result obtained here to previous
analyses using jet rates and differential jet rates. The value obtained for \as\ using a fit
to the \yy\  distribution returns the almost identical result for \as\ as measured with
this analysis. The measurements obtained at higher centre-of-mass energies return smaller uncertainties, 
since the size of the hadronisation correction decreases with increasing centre-of-mass energy. The 
measurement of \as\ using the four-jet rate and data taken between 14 and 44~GeV returns
a smaller hadronisation uncertainty, leading to a decreased overall uncertainty.
\section*{Acknowledgements}
This research was supported by the DFG cluster of excellence "Origin and Structure of the Universe" (www.universe-cluster.de).

\end{document}